\documentclass[prc,aps,preprint,groupedaddress,amsmath,amssymb]{revtex4-1}

\usepackage{bm}

\usepackage{graphicx}

\usepackage{siunitx}
\usepackage{dcolumn}
\usepackage{tabularx}

\begin{document}


\title{$p$-shell hypernuclear structures using the Gogny-interaction shell model}

\author{X. Y. Chen, Z. Zhou, W. G. Jiang, B. S. Hu,}
\author{F. R. Xu}\email{frxu@pku.edu.cn}
\affiliation{School of Physics, and State Key Laboratory of Nuclear Physics and Technology, Peking University, Beijing 100871, China}

\date{\today}

\begin{abstract}
We have systematically investigated the excitation spectra of $p$-shell hypernuclei within the shell model based on the nucleon-nucleon and hyperon-nucleon interactions. For the effective nucleon-nucleon interaction, we adopt the Gogny force instead of the widely-used empirical $p$-shell Cohen-Kurath interaction, while the hyperon-nucleon interaction takes the $\Lambda N$ interaction including the $\Lambda N$-$\Sigma N$ coupling effect. We find that the shell model with the Gogny force can give reasonable descriptions of both spectra and binding energies of the $p$-shell nuclei. With this confidence, combined with the $\Lambda N$ interaction, we have performed shell-model calculations for the {\it p}-shell hypernuclei. We compare our results with $\gamma$-ray data as well as various theoretical calculations, and explain recent experimental hypernuclear excitation spectra observed at JLab.  
\end{abstract}

\pacs{}

\maketitle

\section{Introduction}
Since the first discovery of the $\Lambda$ hypernucleus in 1952 \cite{danysz1953m}, the structures of hypernuclei have been studied in order to understand the dynamics of many-body baryon systems and baryon-baryon interactions. Knowledge about the behaviors of hyperons in finite nuclei and hyperon-nucleon interactions needs to be extrapolated to strange hadronic matter \cite{Gal1995,Schaffner-Bielich2000a} and to neutron stars \cite{Schaffner-Bielich2008,Schaffner-Bielich2010}.

The behavior of a $\Lambda$ hyperon in a nuclear system has been treated using well developed nuclear models such as the shell-model \cite{Gal1971,Millener2010,Millener2012} and mean-field models \cite{Reinhard1989,Sugahara1994a,Shen2006,PhysRevC.76.034312} with an effective $\Lambda$-nucleon interaction. The pioneering calculations of the shell model by Gal {\it et al.}\cite{Gal1971,DALITZ1972109,Gal1978}, have successfully explained the $\gamma$-ray transitions observed in the $p$-shell $\Lambda$ hypernuclei \cite{Dalitz1978,Millener1985} and the $\Lambda$ production cross-sections of the $(K^-,\pi^-)$ and $(\pi^+,K^+)$ reactions \cite{Auerbach1983a,Itonaga1990a}. The effective $\Lambda N$ interaction used in the calculation is based on the Nijmegen potential, and its validity has been well tested in shell-model \cite{Millener1985,Millener2001a} and cluster-model calculations \cite{Motoba1983,Motoba1985,Yamamoto1990a}. More precise calculations need to include the $\Lambda N$-$\Sigma N$ coupling effect which was first suggested by Akaishi and his collaborators \cite{Akaishi2000,10.1007/978-3-7091-6287-3_67} and later tested by {\it ab initio} calculations using the Nijmegen NSC97 potential \cite{Maessen1989,Rijken1999,Rijken2006a}. Shell-model calculations with the inclusion of the $\Lambda N$-$\Sigma N$ coupling have been performed by Millener \cite{Millener2010,Millener2012} in order to interpret the $\gamma$-ray data of the $ N \approx Z $ $p$-shell $\Lambda$ hypernuclei \cite{Tamura2013}. The effective $\Lambda N$ interaction has also been used to investigate the properties of neutron-rich hypernuclei \cite{Umeya2011,Umeya2009,Gal2013}, $\Lambda\Lambda$ hypernuclei \cite{Gal2011} and charge symmetry breaking in $\Lambda$ hypernuclei \cite{Millener2008,Gal2015}.

The effective Cohen-Kurath (CK) nucleon-nucleon ($NN$) interaction \cite{Cohen1965} which was obtained by fitting experimental spectra of nuclei has been widely used in the shell-model calculations of the {\it p}-shell nuclei. While the interaction has successfully explained the data of $ N \approx Z $ $\Lambda$ hypernuclei, calculations for some neutron-rich hypernuclei, such as $^9_\Lambda\text{Li}$, $^{10}_\Lambda\text{Li}$ and $^7_\Lambda\text{He}$, deviates from experimental data. 

In the calculations of shell-model two-body matrix elements (TBMEs), phenomenological potentials have also been used. Delta-type phenomenological interactions, such as surface delta interaction and Skyrme force, have been adopted in shell-model calculations \cite{Green1965a,Sagawa.H1985}. In the present work, we use the finite-range Gogny force to calculate TBMEs and single-particle energies (SPEs). We use the existing Gogny D1S parameters \cite{PhysRevC.21.1568} without refitting. The parameters were determined with mean-field calculations (e.g. Hartree-Fock-Bogoliubov) \cite{PhysRevLett.102.242501,Chappert2008}, giving the good descriptions of the bulk properties of nuclei, such as binding energies, charge radii and saturation densities. Previous shell-model calculations based on the Gogny force have successfully explained both binding energies and spectra of the $p$-shell and {\it sd}-shell nuclei \cite{Jiang2018}, taking the advantage of the density dependence of the Gogny interaction. The present calculations combine the Gogny force and the effective $\Lambda N$ interaction to investigate the {\it p}-shell hypernuclei. The results are compared with those by the Cohen-Kuarth interaction (8-16)TBME \cite{Cohen1965} and recent experiments. 

\section{The framework}
In the present shell-model calculations of the {\it p}-shell hypernuclei, $^4$He is chosen as the frozen core. The $\Lambda$ or $\Sigma$ hyperon is assumed to be in the lowest $0s_{1/2}$ orbit, while $(A-5)$ valence nucleons move in the $0p$ shell. The shell-model TBMEs are calculated using the Gogny force \cite{PhysRevC.21.1568} and the $\Lambda N$ interaction \cite{Gal1978,Millener2010,Gal2016}. The effective Hamiltonian is given by \cite{Millener2010,Umeya2009,Gal2016}
\begin{equation}
	H = H_0 + V_{NN} + V_{YN},
\end{equation}
where $H_0=\Sigma_a {c_a {\hat n}_a}$ is the one-body part of the Hamiltonian in the valence space, with $c_a$ and ${\hat n}_a$ being the energy and particle-number operator for the single-particle orbit $a$, respectively. The $\Lambda$ hyperon is on the $0s_{1/2}$ orbit, and contributes to $H_0$ with a constant single-particle energy, thus does not affect the excitation energy spectrum of the hypernucleus. $V_{NN}$ and $V_{YN}$ are the nucleon-nucleon and hyperon-nucleon interactions, respectively. The hyperon-nucleon interaction includes the direct $\Lambda N$ and $\Sigma N$ interactions and the $\Lambda N$-$\Sigma N$ coupling interaction \cite{Millener2010,Umeya2009,Gal2016}. In the presence of the $\Lambda N$-$\Sigma N$ coupling, the $\Sigma$ hyperon is included in the model space as the excitation state of the $\Lambda$ hyperon. The mass difference between $\Sigma$ and $\Lambda$ hyperons, i.e., the excitation energy of $\Lambda$, is taken as $\SI{80}{MeV}$ \cite{Millener2010,Umeya2009,Gal2016}. 

\subsection{The Gogny $NN$ interaction}
The $NN$ interaction can be written as sum of two-body operators \cite{Brown2006}
\begin{equation}
V_{NN} = \sum\limits_{a\leqslant b,c\leqslant d}\sum\limits_{JT} V_{JT}(ab; cd)\hat{T}_{JT}(ab; cd), 
\end{equation}
with
\begin{equation}
\hat{T}_{JT}(ab; cd) = \sum\limits_{J_zT_z}A_{JJ_zTT_z}^\dagger (ab) A_{JJ_zTT_z}(cd),
\end{equation}
 where $\hat{T}_{JT}(ab; cd)$ is the two-body density operator for the nucleon pair in orbits $(a,b)$ and $(c,d)$ with the coupled angular momentum \(J\) and isospin \(T\). \(A_{JJ_zTT_z}^\dagger\) or \(A_{JJ_zTT_z}\) is the creation or annihilation of the nucleon pair. \(V_{JT}(ab; cd)=\langle a,b|V_{NN,\textbf{Gogny}}|c,d\rangle\) is the antisymmetrized TBME for the $NN$ interaction. As mentioned already, in the present work, we use the finite-range Gogny force \cite{PhysRevC.21.1568},
\begin{equation}
\begin{aligned}
V_{NN,\textbf{Gogny}}=&\sum\limits_{i=1}^2 \mathrm{e}^{-(\bm{r}_1-\bm{r}_2)^2/\mu_i^2}(W_i+B_iP^\sigma-H_iP_\tau- M_iP^\sigma P^\tau)\\
&+t_3\delta(\bm{r}_1-\bm{r}_2)(1+x_0P^\sigma)\left[\rho\left(\frac{\bm{r}_1+\bm{r}_2}{2}\right)\right]^\alpha\\
&+iW_0\delta(\bm{r}_1-\bm{r}_2)(\bm{\sigma}_1+\bm{\sigma}_2)\cdot\bm{k}'\times\bm{k},
\end{aligned}
\label{VNN}
\end{equation}
to evaluate TBMEs. The Gogny force is density dependent. It has been known that the density dependence which originates from the three-body force plays a crucial role in the calculations of nuclear structure and nuclear matter.  In Ref. \cite{Jiang-thesis}, it has been shown that the density dependence is also important for the calculations of nuclear excitation spectra.  In our calculations, the density is determined self-consistently by the numerical iteration with the diagonalizing of  the shell-model Hamiltonian which is density dependent as well, see Ref. \cite{Jiang-thesis} for the detailed explanations of the density iteration and {\it NN} TBME calculations with the Gogny force. The set of most popular parameters, D1S \cite{PhysRevC.21.1568}, is used. 

In Ref. \cite{Jiang-thesis,Jiang2018}, the feasibility to use the Gogny interaction for the effective shell-model $NN$ interaction has been well tested by comparing the TBMEs with those by other effective interactions. Comparisons show good similarity between the TBMEs obtained by the Gogny D1S \cite{PhysRevC.21.1568} and by other empirical or realistic interactions in both $p$ shell and {\it sd} shell \cite{Jiang2018}. 

\subsection{The hyperon-nucleon interaction}
The hyperon-nucleon interaction takes the form as \cite{Gal1978,Millener2010,Gal2016}
\begin{equation}\label{eq:YN}
V_{YN} = \bar{V} +\Delta\bm{s}_N\cdot\bm{s}_{Y} + S_+\bm{l}_{N}\cdot(\bm{s}_N + \bm{s}_{Y}) + S_-\bm{l}_{N}\cdot(-\bm{s}_N + \bm{s}_{Y}) + T S_{12},
\end{equation}
where $Y$ represents the hyperon ($\Lambda$ or $\Sigma$). $\bar{V}$, $\Delta$, $S_+$, $S_-$ and $T$ are radial integrals which are parameterized \cite{Gal1978,Millener2010,Gal2016}. Using Eq.~(\ref{eq:YN}), the $\Lambda N$-$\Sigma N$ coupling can be obtained with different parameter values of the radial integrals \cite{Gal1978,Millener2010,Umeya2009,Gal2016}. $\bm{s}_N$ and $\bm{s}_Y$ are spin operators for the nucleon and hyperon, respectively. $\bm{l}_{N}$ is the angular momentum operator for the nucleon. The tensor operator $S_{12}$ is defined by
\begin{equation}
S_{12}=3(\bm{\sigma}_N\cdot\hat{\bm{r}})(\bm{\sigma}_Y\cdot\hat{\bm{r}})-\bm{\sigma}_N \cdot\bm{\sigma}_Y,
\end{equation}
with $\bm{\sigma} = 2\bm{s}$, and $\hat{\bm{r}} = (\bm{r}_N - \bm{r}_Y)/|\bm{r}_N - \bm{r}_Y|$ is the unit vector of the nucleon-hyperon relative coordinate. We adopt the radial integral parameters determined in Refs. \cite{Millener2010,Umeya2011,Millener2012}.
 
\begin{figure}
\includegraphics[width=0.7\linewidth]{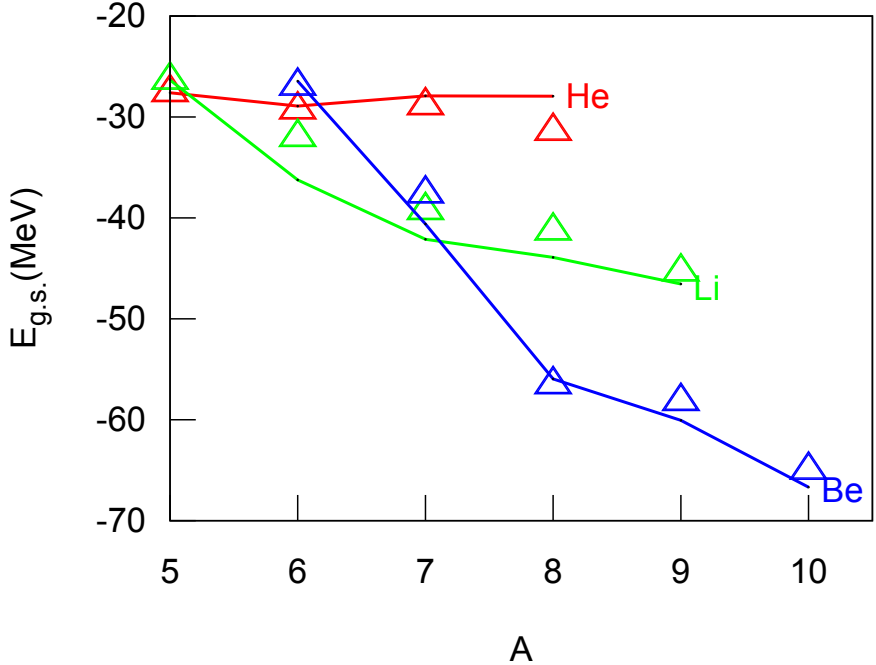}
\caption{Experimental (triangle) and calculated (line) ground-state energies $E_\textbf{g.s.}$ for heliu, lithium and beryllium isotopes. The calculations  were done within the shell-model using the Gogny force. The data are taken from \cite{Tamura2013}.}
\label{fig:bind}
\end{figure}

\section{Calculations of $p$-shell hypernuclei}
As a test how well the shell model based on the Gogny force works, we have calculated the ground-state energies of $A \leq 9$ helium, lithium and beryllium isotopes with the $^4$He core, shown in Fig. \ref{fig:bind}. We see that the calculations are in good agreement with experimental data. In the calculation, for each nucleus, the length parameter $\hbar\omega$ of the harmonic-oscillator (HO) basis takes the value given by minimizing the energy of the nucleus without the $\Lambda$ hyperon. In Ref. \cite{Jiang2018}, we discussed the $\hbar\omega$ choice. The $\hbar\omega$ value determined thus is close to the empirical value of $\hbar\omega=45 A^{-1/3}-25 A^{-2/3}$ \cite{brown2005lecture}. 

As usual, the Coulomb interaction is not included in the shell-model calculation, to keep the isospin symmetry. It has been known the Coulomb effect on the excitation spectrum of a nucleus is small \cite{qi2008isospin}. However, the Coulomb energy needs to be included in the calculation of nuclear binding energy. In Ref. \cite{Jiang2018}, the detailed description of the binding energy calculation has been given, including the calculation of the core energy. It is an advantage of the Gogny-interaction shell model that the core energy can be calculated by the model itself without need of experimental data for the core energy \cite{Jiang2018}. Another advantage of using the Gogny force is that single-particle energies can be obtained with the same TBMEs \cite{Jiang2018} with no need of experimental single-particle energies. Single-particle energies are important inputs in empirical-interaction shell-model calculations \cite{Brown2006}.

\begin{figure*}
		\includegraphics[width=0.3\linewidth]{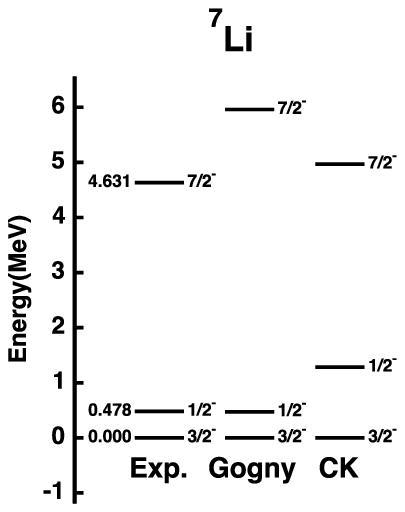} 
		\includegraphics[width=0.3\linewidth]{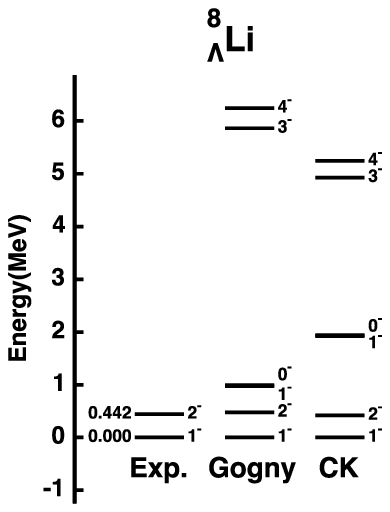} 
		\includegraphics[width=0.3\linewidth]{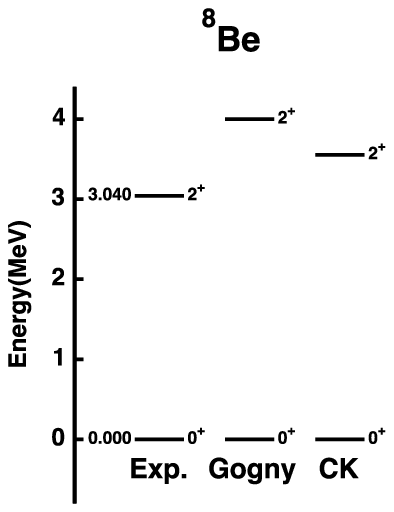} 
		\includegraphics[width=0.3\linewidth]{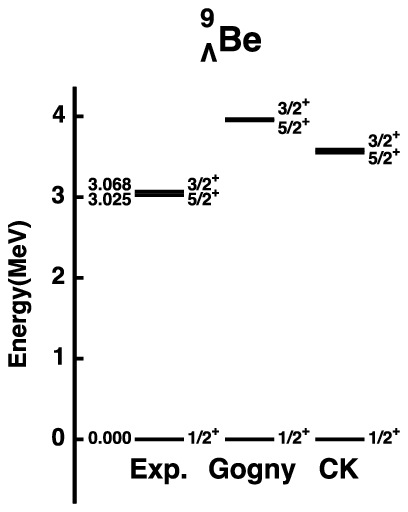} 	
		\includegraphics[width=0.3\linewidth]{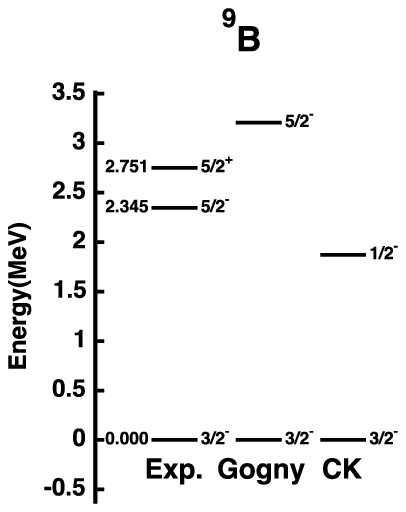} 
		\includegraphics[width=0.3\linewidth]{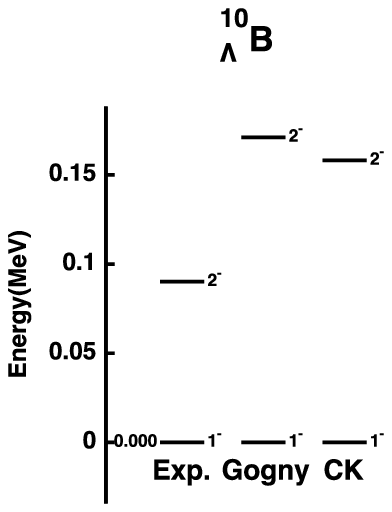}
		\includegraphics[width=0.3\linewidth]{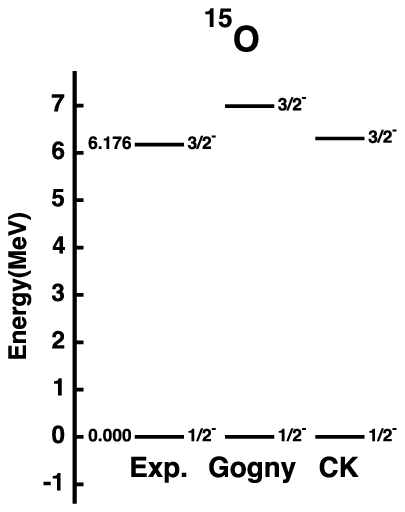}
		\includegraphics[width=0.3\linewidth]{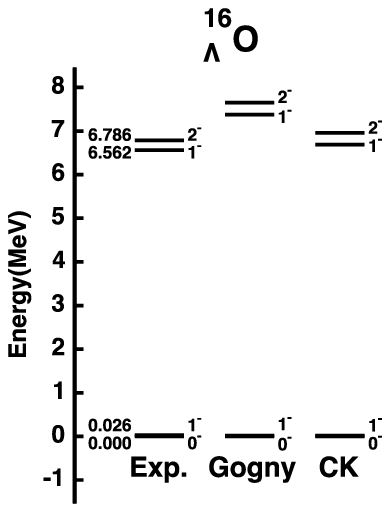} 
	\caption{Spectra for {\it p}-shell hypernuclei $^8_\Lambda \text{Li}$, $^9_\Lambda \text{Be}$, $^{10}_\Lambda \text{B}$ and $^{16}_\Lambda \text{O}$, and adjacent nuclei, calculated by the Gogny+$\Lambda N$ interaction (indicated by Gogny) and the (8-16)TBME(CK)+$\Lambda N$ interaction (indicated by CK), compared with experimental data \cite{Tamura2013}. The experimental spectra of adjacent nuclei are taken from Ref. \cite{Burrows2005}.}
	\label{fig:p1}
\end{figure*}

\begin{figure*}	
		\includegraphics[width=0.3\textwidth]{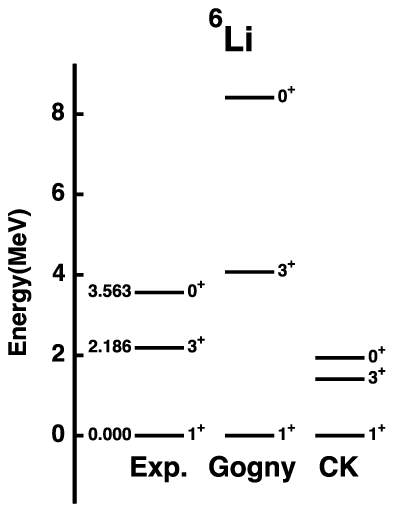} 
		\includegraphics[width=0.3\textwidth]{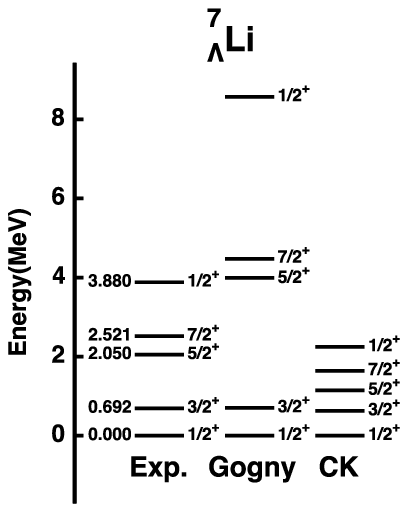} 
		\includegraphics[width=0.3\textwidth]{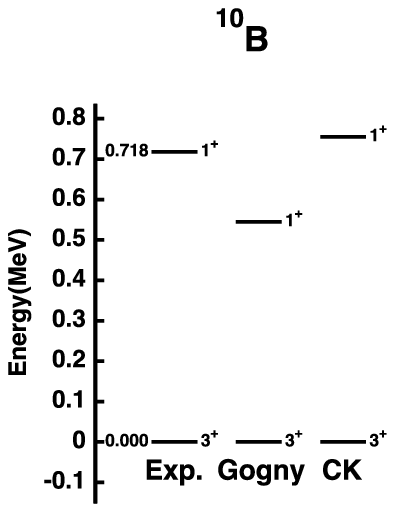} 
		\includegraphics[width=0.3\textwidth]{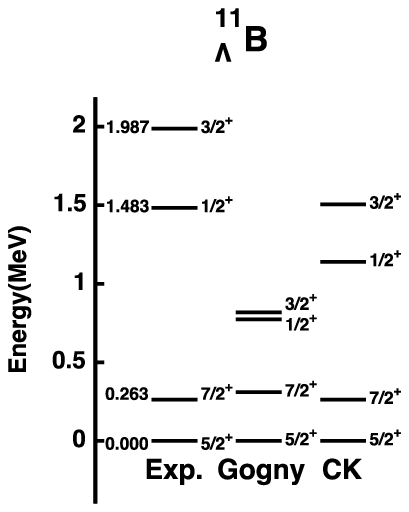}  
		\includegraphics[width=0.3\textwidth]{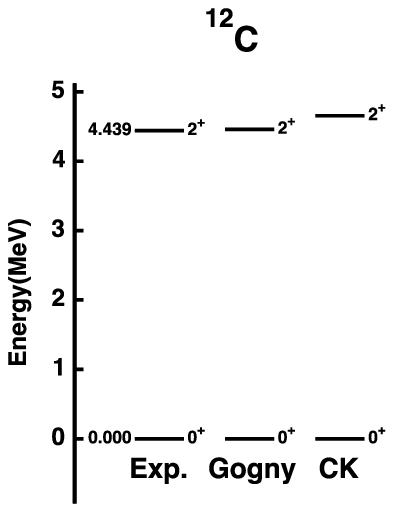}
		\includegraphics[width=0.3\textwidth]{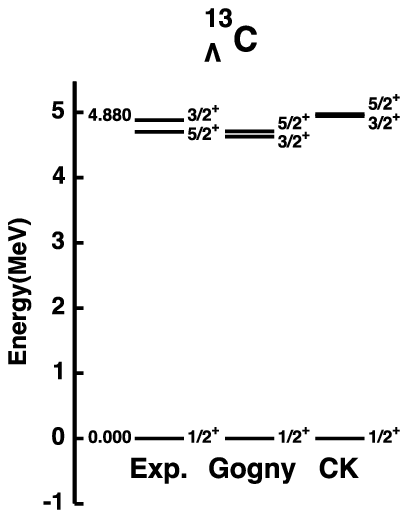} 
		\includegraphics[width=0.3\textwidth]{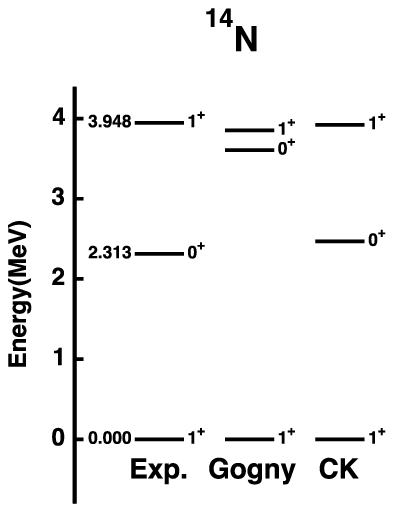} 
		\includegraphics[width=0.3\textwidth]{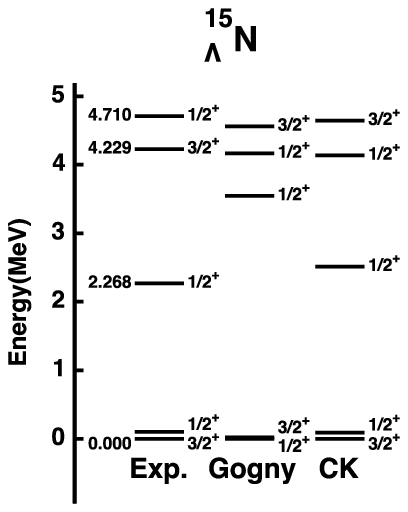}
	\caption{Similar to Fig. \ref{fig:p1}, but for $^7_\Lambda \text{Li}$, $^{11}_\Lambda \text{B}$, $^{12}_\Lambda \text{C}$,  $^{13}_\Lambda \text{C}$, and $^{15}_\Lambda \text{N}$ and their adjacent nuclei. The experimental levels without energy numbering means that their energies have not been resolved exactly in experiments. The data are from \cite{Burrows2005} for nuclei and \cite{Tamura2013} for hypernuclei.}
	\label{fig:p2}
\end{figure*}

With the shell model based on the Gogny $NN$ interaction \cite{PhysRevC.21.1568} and the $\Lambda N$ interaction \cite{Gal1978,Millener2010,Umeya2009,Gal2016}, we have calculated the excitation spectra of the {\it p}-shell hypernuclei in which experimental spectra have been available, shown in Figs. \ref{fig:p1} and \ref{fig:p2}. The spectra of the adjacent nuclei have also been calculated and shown to see how the spectra change with adding a $\Lambda$ hyperon into the nucleus. We see that the Gogny+$\Lambda N$ shell-model calculations reproduce well the observed spectra in both hypernuclei and nuclei in this mass region. The results are also consistent with the calculations with the CK $NN$ interaction \cite{Cohen1965}.

\begin{figure*}
	\includegraphics[width=0.35\textwidth]{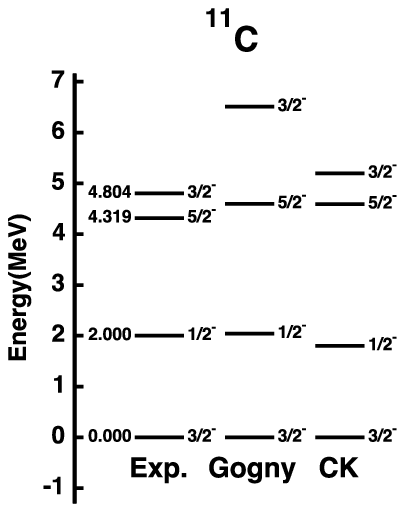}
	\includegraphics[width=0.35\textwidth]{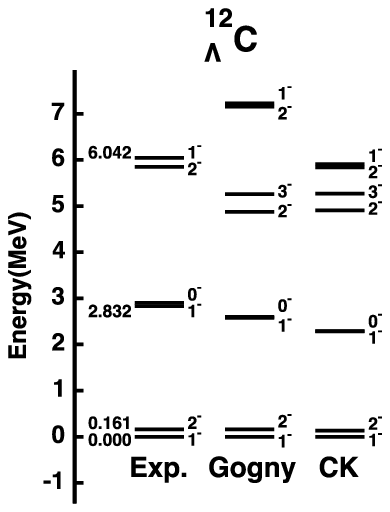}
	\caption{Similar to Figs.~\ref{fig:p1} and \ref{fig:p2}, but for $^{11}$C and $^{12}_\Lambda$C. The experimental data are from \cite{Burrows2005} for $^{11}$C and \cite{Hosomi2015} for $^{12}_\Lambda$C.}
	\label{fig:12C}
\end{figure*}

The level structure of the $^{12}_\Lambda \text{C}$ hypernucleus was determined at Hyperball2 by recent $\gamma$-ray spectroscopy via the $^{12}\text{C}(\pi^+,K^+\gamma)$ reaction \cite{Hosomi2015}. Three experimental excitation levels were determined, corresponding to splitting from the $3/2_1^-$ ground state, $1/2_1^-$ and $3/2_2^-$ excited states in $^{11}$C. As shown in Fig. \ref{fig:12C}, the Gogny force well reproduces the experimental spectrum. The unobserved $3^-$-$2^-$ doublet states with a $\Lambda$ hyperon coupled to the $5/2^-$ state of $^{11}$C are predicted in the present calculation.

\begin{figure*}
	\includegraphics[width=0.36\textwidth]{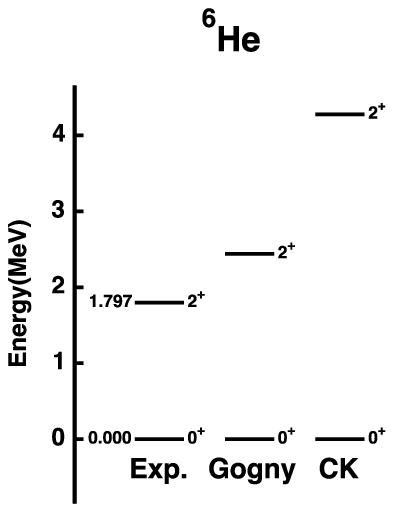} 
	\includegraphics[width=0.6\textwidth]{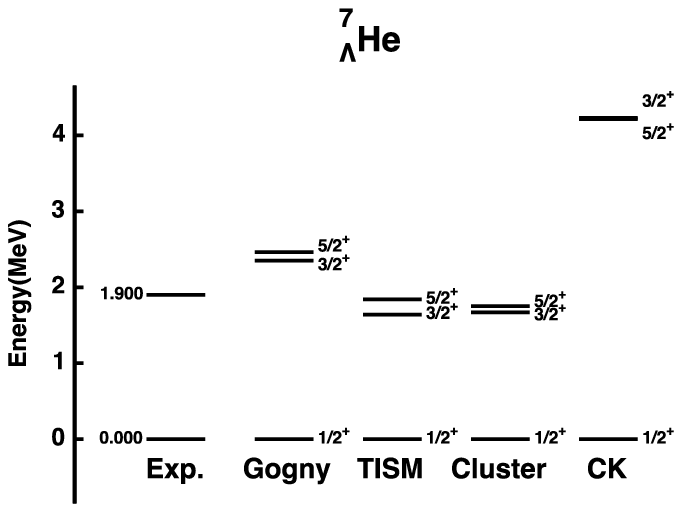} 
	\caption{Similar to Figs.~\ref{fig:p1} and \ref{fig:p2}, but for $^6$He and $^7_\Lambda$He, compared with the calculations by translational invariant shell-model (TISM) \cite{Richter1991}, four-body cluster model (Cluster) \cite{Hiyama2015}. The experimental data are from \cite{Burrows2005} for $^6$He and \cite{Gogami2016a} for $^7_\Lambda$He.}
	\label{fig:he}
\end{figure*}

Recently, the high-resolution spectroscopic experiments for $^7_\Lambda \text{He}$, $^9_\Lambda \text{Li}$, $^{10}_\Lambda \text{Be}$, and $^{12}_\Lambda \text{B}$ have been performed at JLab. The mass spectroscopy of the $^7_\Lambda \text{He}$ hypernucleus was performed using the $^7 \text{Li}(e,e'K^+)^7_\Lambda \text{He}$ reaction \cite{Gogami2016a}. As shown in Fig. \ref{fig:he}, two levels were observed where the lower level was assigned as the ground state and the higher level was assigned as a mixture of $3/2^+$ and $5/2^+$ states, supporting the ``gluelike" behavior of the $\Lambda$ hyperon. In Fig. \ref{fig:he}, we compare our calculations with other theoretical calculations \cite{Richter1991,Hiyama2015}. It is found that the Gogny+$\Lambda N$ model is consistent with those calculated in Refs. \cite{Richter1991,Hiyama2015}.

\begin{figure*}
	\includegraphics[width=0.33\textwidth]{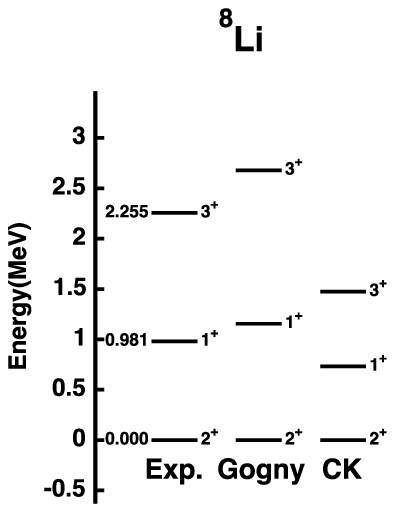} 
	\includegraphics[width=0.5\textwidth]{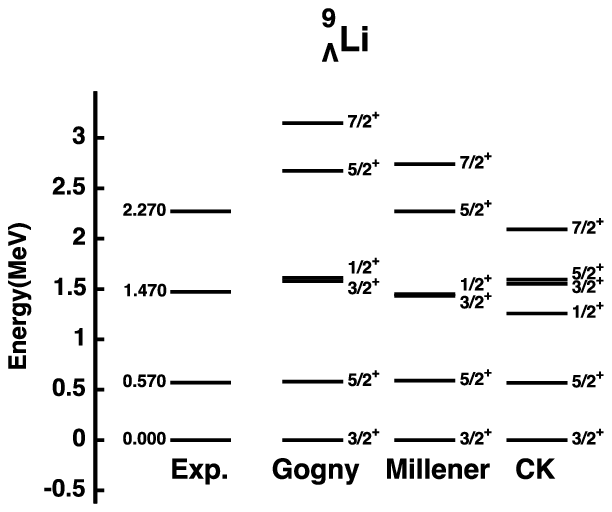} 
	\caption{Similar to Figs. 2-4, but for $^8$Li and $^9_\Lambda$Li. The experimental data are from \cite{Burrows2005} for $^8$Li and \cite{Urciuoli2015} for $^9_\Lambda$Li. ``Millener" indicates the calculation by Millener assuming the $\Lambda$ hyperon coupled to the experimental states of $^8$Li \cite{Millener2012}.}
	\label{fig:Li}
\end{figure*}

The electroproduction of the hypernucleus $^9_\Lambda \text{Li}$ was studied on a $^{9} \text{Be}$ target with sub-MeV energy resolution \cite{Urciuoli2015}. Four levels were obtained by fitting the $^9 \text{Be}(e,e'K^+)^9_\Lambda \text{Li}$ spectrum. Fig.~\ref{fig:Li} shows the experimental level energies \cite{Urciuoli2015} and calculations by the shell-model with the Gongny+$\Lambda N$, CK+$\Lambda N$, and Millener's calculation \cite{Millener2012} which used the $^8 \text{Li}$ spectrum and $\Lambda N$ interaction. The experiment \cite{Urciuoli2015} has not pinned down the spins and parities of the states, but the first two levels were assumed to be the $5/2^+$-$3/2^+$ doublet. The third experimental peak observed in the $^9 \text{Be}(e,e'K^+)^9_\Lambda \text{Li}$ spectrum was assigned as a mixture of $3/2^+$ and $1/2^+$ states with a doublet spacing $< \SI{0.1}{MeV}$ which cannot be resolved with the energy resolution of $\SI{730}{keV}$ in the experiment \cite{Urciuoli2015}. The fourth experimental peak was assumed to be the second $5/2^+$ state \cite{Urciuoli2015}. Our calculation well reproduces the experiment data.

\begin{figure*}
	\includegraphics[width=0.32\textwidth]{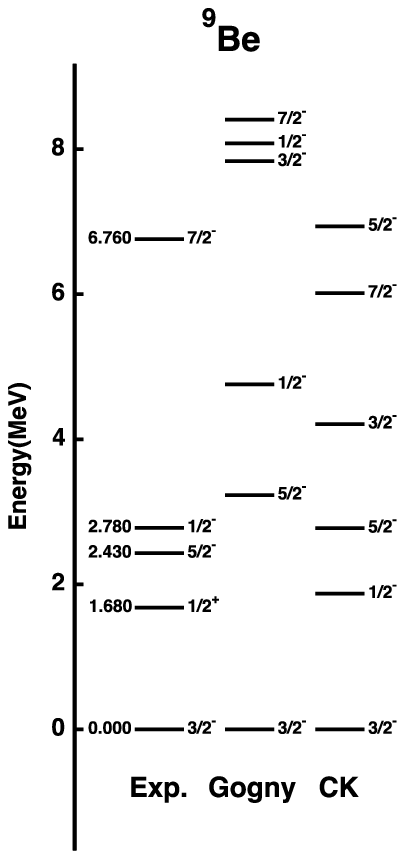} 
	\includegraphics[width=0.65\textwidth]{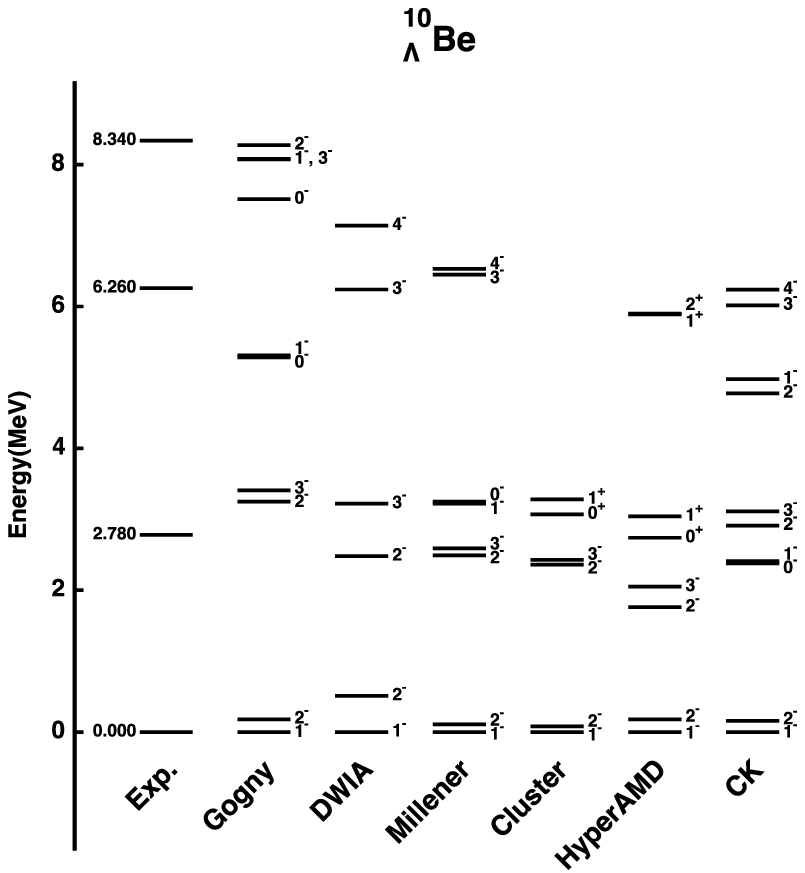} 
	\caption{Similar to Figs. \ref{fig:p1}-\ref{fig:Li}, but for $^9$Be and $^{10}_\Lambda$Be. Other calculations excepting the Gogny and CK are DWIA \cite{Motoba2010}, Millener \cite{Millener2012}, four-body cluster model (indicated by Cluster) \cite{Hiyama2012a}, HyperAMD \cite{Isaka2013}. The data are from \cite{Burrows2005} for $^9$Be and \cite{Gogami2016} for $^{10}_\Lambda$Be. }
	\label{fig:Be}
\end{figure*}

The high-resolution spectroscopy of the hypernucleus $^{10}_\Lambda \text{Be}$ was carried out at JLab, using the $^9 \text{B}(e,e'K^+)^{10}_\Lambda \text{Li}$ reaction, with a resolution of $\sim \SI{0.78}{MeV}$ \cite{Gogami2016}. Four levels were obtained. Fig. \ref{fig:Be} shows the excitation energies of the levels corresponding to the four experimental  peaks \cite{Gogami2016}, compared with various calculations. It is rather convincing to assign the first peak to be the mixture of $1^-$ and $2^-$ states. However, various theoretical calculations disagree with each other in the assignment of the second and third levels. The present calculations reproduce well the experimental spectrum, though higher resolution of spectroscopy is needed to confirm the results.

\begin{figure*}
	\includegraphics[width=0.33\textwidth]{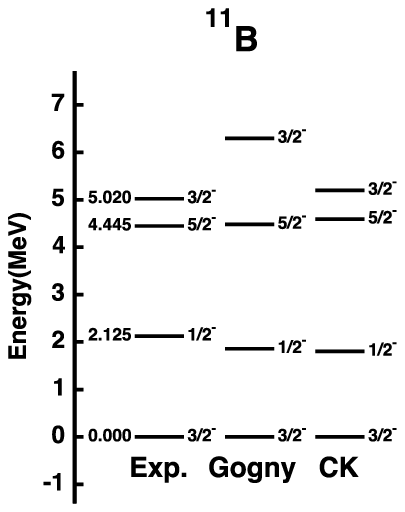} 
	\includegraphics[width=0.33\textwidth]{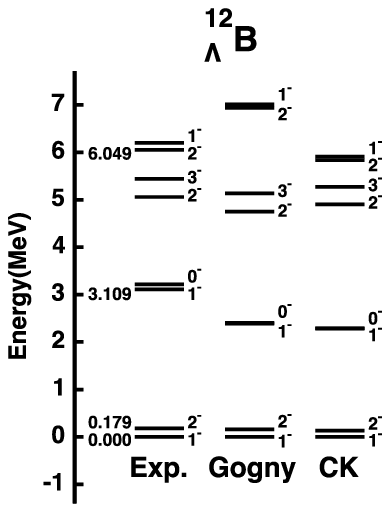} 
	\caption{Similar to Figs. \ref{fig:p1}-\ref{fig:Be}, but for $^{11}$B and $^{12}_\Lambda$B. The data are from \cite{Burrows2005} for $^{11}$B and \cite{Tang2014} for $^{12}_\Lambda$B.}
	\label{fig:B}
\end{figure*}

The spectroscopic experiment for the hypernucleus $^{12}_\Lambda \text{B}$ was carried out at JLab, using the $^{11} \text{C}(e,e'K^+)^{12}_\Lambda \text{B}$ reaction \cite{Tang2014}. Eight peaks in the reaction spectrum were obtained with  statistical significance larger than $4\sigma$. The first four peaks are considered to have a {\it s}-shell $\Lambda$ coupled to $^{11}$B that has a negative-parity structure, while other four peaks correspond to $\Lambda$ in the {\it p}-shell. In this paper, we can only calculate the states with the $\Lambda$ in the {\it s} shell, because we have no $\Lambda N$ interaction for the {\it p}-shell $\Lambda$. The calculations are given in Fig. \ref{fig:B}. For the negative-parity states in which the $\Lambda$ stays in the {\it s} shell, the first two levels are the ground-state doublet states $1^-_1$ and $2^-_1$ with a $\Lambda_s$ coupled to the $3/2^-$ $^{11} \text{B}$ ground state. The third experimental peak is considered to be the lower member of the second doublet ($1^-_2$ and $0^-_1$) with $^{11}$B in the $1/2^-$ configuration. The $0^-_1$ and $1^-_3$, as well as the third doublet ($2^-_2$ and $3^-_1$), were predicted to have small cross sections and thus are difficult to be observed without sufficient statistics and a better signal/background ratio \cite{Tang2014}. Shell-model calculations based on the Gogny or CK interaction reproduce well the experimental spectra of both $^{11}$B and $^{12}_\Lambda$B.

The advantage to use the Gogny force is that we can calculate the spectra of a wide range of nuclei without need to adjust parameters. The density-dependent term provides an equivalent three-body force. In general, the Gogny interaction seems to give better calculations for the neutron-rich {\it p}-shell hypernuclei than the CK interaction.

\section{Conclusion}
We have systematically investigated the excitation spectra of $p$-shell hypernuclei using shell model with the Gogny+$\Lambda N$ interaction including the $\Lambda N$-$\Sigma N$ coupling. With its density-dependence property, the Gogny force can give reasonable descriptions in both spectra and binding energies for a wide range of $p$-shell nuclei without adjusting parameters. For $^8_\Lambda \text{Li}$, $^9_\Lambda \text{Be}$, $^{10}_\Lambda \text{B}$, and $^{16}_\Lambda \text{O}$, the shell-model calculations with the Gogny+$\Lambda N$ interaction gives good agreements with $\gamma$-ray data. For $^7_\Lambda \text{Li}$, $^{11}_\Lambda \text{B}$, $^{12}_\Lambda \text{C}$,  $^{13}_\Lambda \text{C}$, and $^{15}_\Lambda \text{N}$, some of calculated energy spacings deviate slightly from $\gamma$-ray data, but could be improved by further calculations, e.g., by improving single-particle energies and/or increasing model space. With the recent high-resolution spectroscopic experiments for $^7_\Lambda \text{He}$, $^9_\Lambda \text{Li}$, $^{10}_\Lambda \text{Be}$, and $^{12}_\Lambda \text{B}$ performed at JLab, we have compared our calculations with the data as well as other theoretical predictions. Overall agreements have been obtained. We have shown that the Gogny+$\Lambda N$ interaction provides reasonable descriptions for both nuclei and hypernuclei of the {\it p} shell.

\begin{acknowledgments}
This work has been supported by the National Key R\&D Program of China under Grant No. 2018YFA0404401; the National Natural Science Foundation of China under Grants No. 11835001, No. 11575007 and No. 11320101004; the China Postdoctoral Science Foundation under Grant No. 2018M630018, and Beijing innovation research program for undergraduates. We acknowledge the High-performance Computing Platform of Peking University for providing computational resources.
\end{acknowledgments}

\bibliography{refnucl}

\end{document}